\documentstyle[11pt]{article}
\titlepage
\newcommand{\be}{\begin{equation}}

\newcommand{\ee}{\end{equation}}

\newcommand{\bea}{\begin{eqnarray}}
\newcommand{\eea}{\end{eqnarray}}

\begin{document}


\title
{ Cosmological Vacuum in Unified Theories}

\author{
V.N.Pervushin, V.I Smirichinski \\[0.3cm]
{\normalsize\it Joint Institute for Nuclear Research},\\
 {\normalsize\it 141980, Dubna, Russia.}}

\date{\empty}

\maketitle
\medskip


\begin{abstract}
{\large
{ The unification of the Einstein theory of gravity with a
conformal invariant version of the standard model for electroweak interaction
without the Higgs potential is considered. In this theory, a module of the
Higgs field is absorbed by the scale factor component of metric so that the
evolution of the Universe and the elementary particle masses have one and the
same cosmological origin and the flat space limit corresponds to the
$\sigma$-model version of the standard model.
The red shift formula and Hubble law are obtained under
the assumption of homogeneous matter distribution.
We show that the considered theory leads to a very small vacuum density of the
Higgs field $\rho_\phi^{Cosmic}=10^{-34}\rho_{cr}$ in contrast with
the theory with the Higgs potential $\rho_\phi^{Higgs}=10^{54}\rho_{cr}$.
}}

\end{abstract}

\section{Introduction}
The homogeneous scalar field, generating elementary particle masses in the
standard model (SM) for electroweak interactions, is based on the Higgs
potential. The physical motivation for this potential as a consequence
of the first symmetry principles is unclear and there are a number of
difficulties in both cosmology (great vacuum density \cite{1}, monopole
creation \cite{2}, domain walls \cite{3}) and the standard model
(breakdown of perturbation theory for the expected values of the Higgs
mass $m_H\sim 1 TeV$ ~\cite{4,5,6}).
The present talk is devoted to a unification of Einstein's theory of gravity
with the conformally invariant version of the standard model for electroweak
interactions (without the Higgs potential) and to the cosmological
consequences of this theory.
\section{Model}
We consider Einstein's theory supplemented by the conformal invariant
part of SM. The conformal symmetry principle switches off the Higgs
potential and introduces the Penrose-Chernikov-Tagirov term \cite{10} so
that the initial action for our consideration consists of two parts
\be     \label{T}
W_{tot} = W_{SM}^c(g,\phi,W,Z,A,e,\nu) + W_{E}(g,\phi).
\ee
The first part  $W_{SM}^c$ represents the conformal invariant part of SM
and it depends on the entirety of fields with conformal
weights $(n)$.
The second part of the action (\ref{T}) $W_{E}(g,\phi)$ represents the
Einstein-Hilbert action with the conformally coupled scalar field \cite{10}
\be     \label{E}
W_{E}(g,\phi)=\int dtdx^3 \sqrt{-g}\left[ -\frac{^{(4)}R(g)}{6}
(\mu^2-\left| \phi \right| ^2)+\partial_\mu \left| \phi
\right| \partial ^\mu \left| \phi \right| \right];~~(\mu^2=\frac{3M_{Pl} ^2}{8\pi}).
\ee
We express the total action (\ref{T}) in terms of the conformal invariant
variables (${}^{(n)}F_c ={}^{(n)}Fa^{-n}$) extracting the space-scale
factor $a= [{}^{(3)}g]^{1/6};~~(g_{\mu\nu}= a^2 g^{(c)}_{\mu\nu}~)$.
This choice of variables can be justified by the principle of
the conformal invariance of dynamical variables \cite{8}.

The action $ W_{SM}^c $ does not depend on the scale variable $ a $
due to the conformal invariance.
The action (\ref{E}) $W_{E}$ has the symmetric form  with respect to the
scale factor $\bar a= \mu a$
and the scalar field $ \phi_c $ with $ \sqrt{-g^{(c)}} = N_c $
\be     \label{EC}
W_{E} =\int dx^4 N_c\left[ -\frac{^{(4)}R(g^{(c)})}{6}
(\bar a^2- \phi_c  ^2)+
\partial_\mu  \phi_c  \partial^\mu  \phi_c  -\\
\partial_\mu \bar a \partial^\mu \bar a \right].
\ee
\section{Absorption of the Higgs field by the space metric}
The class of physical solutions for the theory (\ref{EC}) should be restricted
by the condition
\be     \label{R}
(\bar a^2- \phi_c  ^2)\geq 0,
\ee
in the whole four-dimensional space. In the opposite case, the Einstein action
changes sign with respect to the matter one, and gravitation converts into
antigravitation with a wrong sign of the Newton interaction and negative energy
for gravitons. The rough analogy of this restriction is the light
cone in special relativity which defines the physically admissible region of
the particle motion.
The restriction (\ref{R}) leads to the symmetric initial data
$\bar a=0, \phi_c=0$.
The symmetric equations (action) and symmetric initial data can lead only
to a symmetric class of solutions of the equations of motion and constraints
$ \phi_c ={\pm} \bar a $.
This solution can be treated as dynamical absorption of
the conformal scalar field by the scale factor component of the metric.
 In terms of the initial scalar field module  $|\phi|=\phi_c/a$
we have only the vacuum value
$|\phi|= \mu$,
in contrast with the decomposition of the Higgs field in the potential
model.
We got accustomed to the decomposition of a scalar field over plane waves
treating them as particle-like excitations. In fact, a correct
decomposition includes (in addition to plane waves) the zero-mode sector.
Here, we face the case when the scalar field  looses its particle-like
excitations and has only the zero-mode sector formed
by the scale factor.
\section{Cosmology}
In the vicinity of the beginning of the Universe $\phi_c ={\pm} \bar a=0$
the total action (\ref{T})  describes only a set of  SM massless
fields (i.e. radiation).
At first, we restrict our consideration to the harmonic excitations of
these fields, in the FRW metric.
The consistent classical and quantum descriptions of the Universe
filled in by these massless harmonic excitations implies two stages:
dynamical (D) and geometrical (G) \cite{8}.\\
(D) At the dynamical stage, parameters of the time-reparametrization
transformations are completely separated from the sector of
physical variables as a result of which the scale factor $a$ converts
into the invariant conformal time of evolution of the Universe \cite{8}
and leaves the set of independent physical variables of the theory.\\
G) The geometrical stage is the transition to the Friedmann comoving frame of
reference connected with the massive dust. The Friedmann  observables, in
the comoving frame of reference, are constructed by conformal transformation
of the dynamical (conformal)
variables and coordinates including the Friedmann time interval
and distance \cite{8} $dt_F=a d\eta; D_F=a D_c $.

In the case considered, the Einstein - Friedmann equation
($ \delta W^H / \delta N_c^0=0 $) represents the sum of energy densities
 of the scale factor (cr), scalar field ($\phi$), and "radiation" (R)
 $$  - \rho_{cr} + \rho_{\phi}^0 + \rho_{R} = 0.$$
The evolution of the cosmic scale $a(t_F)$  coincides
with the one of the Friedmann Universe filled by radiation
The scalar field $\phi_c$ repeats  this evolution,
while the initial scalar field $|\phi|=\frac{\phi_c}{a}$
is equal to a constant
$|\phi|= \mu ({\rho_\phi^0}/{\rho_{cr}})^{\frac{1}{2}}.$
The value of this scalar-field, which follows from the Weinberg-Salam theory
$\sqrt{g^2/2} |\phi| = m_W \sim 10^2 GeV$ ,
allows us to estimate the value of the relation of energy
densities of the scalar field $(\rho_{\phi}^0)$ and the expansion of the
Universe $(\rho_{cr})$:
$\rho_\phi^{Cosmic}\sim 10^{-34}\rho_{cr}$.
Recall that the Higgs potential leads to the opposite situation
(see \cite{1})
$\rho_\phi^{Higgs}\sim 10^{54}\rho_{cr}$.

When masses of the SM elementary particles (determined by $\phi_c(\eta)$)
become greater than their momenta, we should include additional terms
in the energy density of the scalar field of the type of
$ \rho_\phi=\rho_\phi^0-\phi_c<n_f> +\phi_c^2 <n_b^2>$
associated with the fermion and boson "dusts" at rest.
Here $\rho_\phi,<n_f> and <n_b^2>$ are phenomenological parameters which
determine the solution to the homogeneous scalar field equations.
For the  case considered, we have obtained the oscillator - like solution for
the conformal scalar field \cite{4}
\be\label{s1}
\phi_c(\eta) = \rho_\phi^{1/2}{\omega_\phi}^{-1}\sin\omega_\phi\eta+\frac{1}{2}<n_f>\omega_\phi^{-2}
(1-\cos \omega_\phi\eta),~~~(\omega^2_\phi=1/r_0^2+<n_b^2>).
\ee
If the dust term dominates,
the SM-particle masses ($\phi_c/a$) become  dependent on time.
A photon radiated by an atom on an astronomical object (with a distance $D$
to the Earth) at the time $t_F -D$ remembers the value of this mass at
this time.
As a result, the red shift is defined by the product of two
factors: the expansion of the Universe space $(a)$  and the alteration of the
elementary particle masses $(\phi_c/a)$.
Finally, we get the red shift $Z$ and the Hubble law
\be \label{cc}
Z(D)=\frac{\phi_c(t_F)}{\phi_c(t_F-D)}-1\;\;;\;\;H_0=\frac{d\phi_c}{\phi_cdt_F}.
\ee
Thus, in the theory considered, one and the same function $(\phi_c)$  describes
both the expansion of the Universe and masses of elementary particles.
In the flat-space limit $^{(4)}R(g)=0$, we get the $\sigma$-model version
of SM without Higgs particles, which is discussed now in order to remove
 the difficulty of large coupling constant
in the Higgs sector \cite{5,6}.

\end{document}